\documentclass[aps,prb,twocolumn,showpacs,10pt]{revtex4}
\usepackage[utf8]{inputenc}
\usepackage{graphicx,amsmath}
\usepackage{amssymb}
\usepackage{amsthm}

\begin{document}
\title{Topological invariant for generic 1D time reversal symmetric superconductors in class DIII}
\author{Jan Carl Budich, Eddy Ardonne}

\affiliation{Department of Physics, Stockholm University, SE-106 91 Stockholm, Sweden}
\date{\today}
\begin{abstract}
A one dimensional time reversal symmetric topological superconductor (symmetry class DIII) features a single Kramers pair of Majorana bound states at each of its ends. These holographic quasiparticles are non-Abelian anyons that obey Ising-type braiding statistics. In the special case where an additional $U(1)$~spin rotation symmetry is present, this state can be understood as two copies of a Majorana wire in symmetry class D, one copy for each spin block. We present a manifestly gauge invariant construction of the topological invariant for the generic case, i.e.,  in the absence of any additional symmetries like spin rotation symmetry. Furthermore, we show how the presence of inversion symmetry simplifies the calculation of the topological invariant. The proposed scheme is suitable for the classification of both interacting and disordered systems and allows for a straightforward numerical evaluation of the invariant since it does not rely on fixing a continuous phase relation between Bloch functions. Finally, we apply our method to compute the topological phase diagram of a Rashba wire with competing s-wave and p-wave superconducting pairing terms.      
\end{abstract}
\maketitle

\section{Introduction}
Triggered by the discovery of the one dimensional topological superconductor (1DTSC) by Kitaev in 2001 \cite{Kitaev2001}, one dimensional topologically nontrivial proximity induced superconductors have received enormous attention in recent years \cite{HasanKane,XLReview2010,BeenakkerMajReview,AliceaReview}. The theoretical prediction of the 1DTSC phase in nanowires coupled to a conventional superconductor \cite{SauTSC,OppenTSC} has paved the way for first experimental signatures of Majorana bound states (MBS) in condensed matter physics \cite{LeoMaj, LarssonXu, HeiblumMaj}. The non-Abelian statistics of these exotic quasiparticles have been theoretically demonstrated \cite{AliceaBraiding} with the help of T-junctions of wires.
\\

\begin{figure}[b]
\includegraphics[width=0.60\columnwidth]{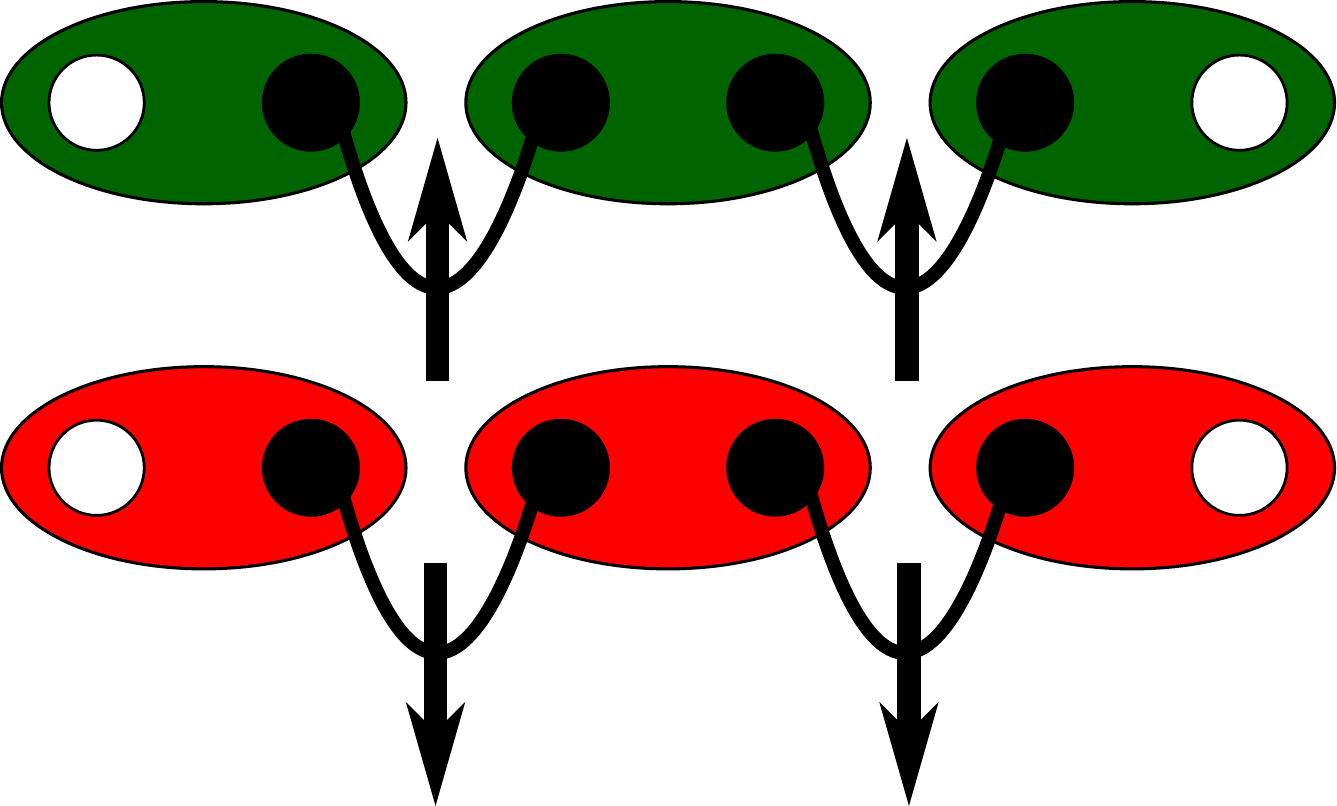}
\caption{Schematic representation of a 1DTSC in class DIII, consisting of two Kramers partners (red and green) of Kitaev's Majorana chain in class D. The white dots denote the Majorana bound states at the ends. The ovals denote the lattice sites hosting two paired Majorans (connected black dots). The arrows denote the localization of the occupied quasiparticles illustrating the polarization of the chain with respect to the oval lattice sites.}
\label{fig:d31dphase}
\end{figure}

While non symmetry protected topological superconductors (TSC)s belong to the symmetry class D \cite{AltlandZirnbauer}, there are also time reversal symmetry (TRS) protected TSCs in symmetry class DIII in one, two, and three spatial dimensions \cite{RoyTSC,Schnyder2008,QiTSC}. Due to Kramers theorem, the TRS preserving 1DTSC features a Kramers pair of spinful MBS at each of its ends, as depicted schematically in Fig.~\ref{fig:d31dphase}.
Generically, pairs of MBS are topologically equivalent to ordinary fermions. Hence, one might at first glance expect that these end states obey ordinary fermionic braiding statistics. However, it has recently been shown \cite{LawD3Braiding} that this is not the case for single Kramers pairs of MBS: As long as TRS is preserved each of the MBS Kramers partners obeys Ising anyonic statistics independently. Several authors have recently proposed realizations of the TRS preserving 1DTSC in class DIII \cite{Law1DD3,Nakosai1DD3,KeselmanDIII2013} (see also \cite{Motrunich2001}). Given the fact that Ising anyons like MBS do not allow for universal topological quantum computing per se \cite{NayakReview}, the additional Kramers (spin) degree of freedom might be considered helpful for quantum information processing architectures based on MBS.\\

By taking a look into the periodic table of topological states of matter \cite{Schnyder2008,KitaevPeriodic,RyuLudwig}, we infer that a $\mathbb Z_2$~invariant can be assigned to the symmetry class DIII in 1D. Here, we are concerned both with the physical meaning and the practical calculation of the value of this invariant $\nu = \pm 1$.  In the presence of an additional $U(1)$~ spin rotation symmetry that fixes a global spin quantization axis, the TRS preserving 1DTSC can be understood as two copies of the non symmetry protected 1DTSC, each copy representing one spin projection. In this special case, the calculation of $\nu$~boils down to the calculation of Kitaev's Pfaffian invariant \cite{Kitaev2001} for one of the spin blocks. In the generic case without additional symmetries, the situation becomes more complicated. The calculation of $\nu$~as originally proposed in Ref. \onlinecite{RyuLudwig} then involves a twofold dimensional extension to connect the system to its parent state, the 3D TSC in class DIII. Such a calculation can be pretty cumbersome to evaluate in particular numerically. For non-interacting systems with closed boundary conditions, a scattering matrix approach as reported in Ref. \onlinecite{FulgaSMatrixTI} can be used.\\

\section{Main results}
In this work, we prove that the topological $\mathbb Z_2$~invariant $\nu$~defining the 1DTSC in class DIII can be viewed as a Kramers polarization. In the presence of a fixed spin quantization axis this Kramers polarization reduces to the well known Pfaffian invariant \cite{Kitaev2001} for one spin block. Most interestingly, we provide a manifestly gauge invariant way to directly calculate this bulk invariant for generic systems in the absence of any additional symmetries. By manifestly gauge invariant, we mean that no continuous phase relation between wave functions at different $k$-points needs to be known, which allows for a straightforward numerical calculation of $\nu$. As has been shown previously for topological insulators in the symplectic class AII \cite{FuInversion2007}, we find a tremendous simplification regarding the analytical form of $\nu$~in the presence of inversion symmetry. The calculation of $\nu$~then only involves the representation of the parity operator at the real $k$-points $0$~and $\pi$, where $k=-k$. By rephrasing $\nu$~in terms of the single particle Green's function and applying twisted boundary conditions, the definition of the invariant can readily be extended to interacting and disordered systems, respectively (see also \cite{TurnerEnt2011,Fidkowski2011} for invariants for interacting systems).\\

\section{Manifestly gauge invariant construction of the invariant}
We would like to construct a generally valid, manifestly gauge invariant, and physically intuitive formulation of the topological invariant $\nu$~characterizing the 1DTSC in symmetry class DIII \cite{AltlandZirnbauer}. A Bogoliubov de Gennes (BdG) Hamiltonian $H$~in this symmetry class is characterized by the presence of two antiunitary symmetries. A symplectic TRS $\mathcal T$~satisfying $\mathcal T^2=-1$~and a particle hole symmetry (PHS) $\mathcal C$~satisfying $\mathcal C^2=1$. $H$~commutes with $\mathcal T$~and anticommutes with $\mathcal C$. Consequently, the combination $U_{\text{CS}}= \mathcal T\circ \mathcal C$~is a unitary operation that anticommutes with the Hamiltonian.  $U_{\text{CS}}$~is called a chiral symmetry.  For concreteness, we will without loss of generality use the common convention $\mathcal T=i\sigma_yK$~and $\mathcal C=\tau_x K$, where $\sigma_y$~is a Pauli matrix in spin space whereas $\tau_x$~is a Pauli matrix in Nambu space and $K$~denotes complex conjugation. We would like to point out that the PHS is not a physical symmetry but rather results trivially from the redundancy of the BdG description of superconductivity which deals with two copies of the physical spectrum: A particle and a hole copy. \\

Let us start with the special case where an additional $U(1)$-spin rotation symmetry gives rise to a global spin quantization axis: We assume that $\left[H,\sigma_z\right]=0$. It follows immediately that the Hamiltonian can be block diagonalized into spin blocks as $H=\text{diag}\left(h,h^*\right)$, where the diagonal matrix structure is in spin space. The individual blocks are in symmetry class D and are characterized by the same value $\mathcal M=\pm 1$~of Kitaev's Pfaffian $\mathbb Z_2$~invariant \cite{Kitaev2001}. For this special case, the topological invariant of the full system $\nu$~we are concerned with in this work is hence just given by the invariant for one of its spin blocks. The physical interpretation of this phenomenology is straightforward. In the non-trivial phase, each of the spin blocks features a single spin polarized MBS at each end of the system, i.e., there is a spin-degenerate pair of MBS associated with each end. In the trivial phase, each spin block has zero (or at least an even number of) MBS. Equivalently to the Pfaffian invariant  $\mathcal M$, the 1DTSC can also be characterized by a Zak-Berry phase \cite{Berry,ZakPol} which is quantized to integer multiples of $\pi$~due to the presence of PHS \cite{HatsugaiQuantizedBerry}. The quantized Zak-Berry phase can be understood as a half-integer quantized polarization of the BdG band structure \cite{ZakPol}. We hence have
\begin{align}
\nu (H)=\mathcal M(h)=\exp\left(i\int_0^{2\pi}\text{d}k\,\mathcal A_o^\sigma(k)\right)=\pm 1, 
\end{align}
where $\mathcal A^\sigma_o(k)=-i \sum_{\alpha:\, {\rm occ}}\langle u_\alpha^\sigma(k)\rvert \partial_k \lvert u_\alpha^\sigma(k)\rangle$~is the Berry connection of the spin block $\sigma$~and the sum on $\alpha$~runs over its occupied bands with Bloch states $\lvert u_\alpha^\sigma(k)\rangle$. As already mentioned the polarization $P_o^\sigma=\frac{1}{2\pi}\int_0^{2\pi}\text{d}k\,\mathcal A_o^\sigma(k)$~does not depend on the spin index $\sigma$~modulo integers.\\

We now turn to generic Hamiltonians in class DIII and hence drop the assumption of a $U(1)$~spin rotation symmetry. However, the presence of TRS still implies that the eigenstates come in pairs. Both members of such a pair have degenerate energies at the time reversal invariant (i.e., real) momenta due to Kramers theorem. Instead of a spin index $\sigma=\uparrow,\downarrow$, the Bloch bands can hence be assigned a Kramers index $\kappa = I,II$. Following the general analysis of Bloch functions in the presence of a symplectic TRS in Ref. \onlinecite{FuPump} we define
\begin{align}
&\lvert u_\alpha^I(-k)\rangle=-\text{e}^{i \chi_\alpha(k)} \mathcal T\lvert u_\alpha^{II}(k)\rangle\nonumber\\
&\lvert u_\alpha^{II}(-k)\rangle=\text{e}^{i \chi_\alpha(-k)} \mathcal T\lvert u_\alpha^{I}(k)\rangle.
\label{eqn:trsBloch}
\end{align}
This conjugation property of the Kramers bands leads to the constraint on the Berry connection of the Kramers blocks \cite{FuPump}
\begin{align}
\mathcal A_o^I(-k)=\mathcal A_o^{II}(k)-\sum_{\alpha:\, {\rm occ}} \partial_k \chi_\alpha(k),
\label{eqn:trsBerry}
\end{align}
i.e., the Berry connections of opposite Kramers blocks at opposite momenta are related by a gauge transformation. Eq. (\ref{eqn:trsBerry}) implies that the associated Kramers polarizations $P_o^\kappa=\frac{1}{2\pi}\int_0^{2\pi}\text{d}k\,\mathcal A_o^\kappa(k)$~are independent of $\kappa$~modulo integers. This generalizes our previous statement that the Kitaev invariant $\mathcal M$~is the same for both spin blocks in the $\sigma_z$~conserving case to the generic case of Kramers blocks. This clearly shows that the Kitaev invariant for the total TRS preserving Hamiltonian is always trivial as it consists of two identical contributions from the Kramers blocks. This is consistent with the fact that only Kramers pairs of MBS can occur at the ends of a TRS preserving 1DTSC as opposed to the single MBS in the nontrivial TRS breaking 1DTSC. Using Eqs. (\ref{eqn:trsBloch}),(\ref{eqn:trsBerry}), the Kramers polarization $P_o^I$~can be readily expressed as \cite{FuPump}
\begin{align}
P_o^I=\frac{1}{2 \pi}\left[\int_0^\pi \text{d}k \mathcal A_o(k)~+~i\log\left(\frac{\text{Pf}~\theta_o(\pi)}{\text{Pf}~\theta_o(0)}\right)\right],
\label{eqn:fukanePol}
\end{align}
where $\mathcal A_o (k)=\mathcal A_o^I (k)+\mathcal A_o^{II} (k)$ and Pf denotes the Pfaffian. The matrix form of $\mathcal T$ is denoted by $\theta(k)$~which is antisymmetric at the real $k$-points $k=0,\pi$; $\theta_o(k)$ denotes the restriction of $\theta(k)$ to the occupied bands. Ref. \onlinecite{FuPump} is concerned with the symplectic symmetry class AII which only requires TRS. For generic Hamiltonians in AII, $P_o^I$~is not quantized which is reflected in the fact that there are no topologically non-trivial states in this class in 1D \cite{Schnyder2008,KitaevPeriodic}. In symmetry class DIII however, the additional presence of the spectrum generating PHS implies that the polarization is half-integer quantized even for the individual Kramers blocks. Hence, the value of $P_o^I (\text{mod}~1)$~defines a $\mathbb Z_2$~invariant in class DIII.\\

Several remarks on Eq. (\ref{eqn:fukanePol}) are in order. It has already been pointed out in Ref. \onlinecite{FuPump} that the expression on the right hand side of Eq. (\ref{eqn:fukanePol}) is gauge invariant. However, its calculation requires the fixing of an arbitrary gauge for which a continuous phase relation between the Bloch functions in half of the Brillouin zone has to be known. Hence,  Eq. (\ref{eqn:fukanePol}) does not yet provide a constructive prescription as to the numerical calculation of the topological invariant. We will now proceed to construct such a manifestly gauge invariant recipe. Our construction makes use of the manifestly gauge invariant formulation of the the adiabatic theorem due to Kato \cite{Kato1950} which works with projection operators rather than wave functions. To this end, we first exponentiate Eq. (\ref{eqn:fukanePol}),
\begin{align}
\nu=\text{e}^{i2\pi P_o^I}=\text{e}^{i\int_0^\pi \text{d}k \mathcal A_o(k)}  ~ \Bigl(\tfrac{\text{Pf}~\theta_o(0)}{\text{Pf}~\theta_o(\pi)} \Bigr) =\pm 1.
\label{eqn:nuresult}
\end{align}
The Kato connection associated with the occupied bands is defined as \cite{Kato1950,AvronQuadrupole1989,TSMReview}
\begin{align}
\mathcal A_o^K(k)=-\left[(\partial_k \mathcal P_o(k)),\mathcal P_o(k)\right], 
\end{align}
where $\mathcal P_o(k)$~denotes the basis independent projector onto the occupied bands.
In Ref. \onlinecite{TSMReview}, it has been demonstrated that the propagator associated with the full non-Abelian Berry connection is nothing but the matrix representation of the Kato propagator $\mathcal U^K=T\text{e}^{-\int \mathcal A_o^K}$~associated with the Kato connection $\mathcal A_o^K$. The Abelian part of this propagator is then simply given by the determinant of this unitary representation matrix. Remarkably, the Kato propagator can be calculated numerically in a straightforward way in contrast to the Berry connection. Explicitly, for the path $0\rightarrow \pi$~in $k$-space appearing in Eq. (\ref{eqn:nuresult}), we get (see Refs. \cite{AvronQuadrupole1989,TSMReview} for the general construction)
\begin{align}
\mathcal U^K(0,\pi)=\lim_{n\rightarrow \infty}\Pi_{j=0}^n  \mathcal P_o(k_j),\quad k_j=j\frac{\pi}{n},
\label{eqn:katoprop}
\end{align}
where the product is ordered from the right to the left with increasing $j$. The practical calculation of this quantity only requires the calculation of the gauge-independent projectors $\mathcal P_o (k)$~onto the occupied bands on a discrete mesh of points in $k$-space. To proceed with the evaluation of the invariant $\nu$~as defined in Eq. (\ref{eqn:nuresult}), we only have to fix an arbitrary basis of occupied Bloch functions $\left\{\lvert \alpha\rangle\right\}_\alpha$~at $k=0$~and $\left\{\lvert\tilde\alpha\rangle\right\}_\alpha$~at $k=\pi$. Note that this choice does not require any information about relative phases of Bloch functions at different points in $k$-space. Instead we are allowed to pick an arbitrary basis at each of the points $k=0$~and $k=\pi$. We define the matrix representation of the Kato propagator in this basis choice as $\hat U^K_{\alpha,\beta}=\langle \tilde \alpha\rvert \mathcal U^K(0,\pi)\lvert \beta\rangle$. The representation matrices of $\mathcal T$~are denoted by $\bigl(\hat \theta_o (0)\bigr)_{\alpha\beta}=\langle \alpha\rvert \mathcal T\lvert \beta\rangle$~and $\bigl(\hat \theta_o(\pi)\bigr)_{\alpha\beta}=\langle \tilde\alpha\rvert \mathcal T\lvert \tilde\beta\rangle$, respectively. With these definitions Eq. (\ref{eqn:nuresult}) can be simplified to
\begin{align}
\nu=\left(\text{det}~\hat U^K\right)\frac{\text{Pf}~\hat \theta_o(0)}{\text{Pf}~\hat \theta_o(\pi)}=\pm 1,
\label{eqn:nusimple}
\end{align}
where $\nu=-1$~defines the topologically non-trivial phase. Eq. (\ref{eqn:nusimple}) is the key result of the present work. It allows an even numerically straightforward prescription to calculate the topological $\mathbb Z_2$~invariant of a generic 1DTSC in symmetry class DIII. In an example below, we show that this invariant does indeed distinguish between the trivial and non-trivial 1DTSCs in class DIII.\\

\section{Competition of s-wave and p-wave pairing in a Rashba wire}
\begin{figure}[t]
\includegraphics[width=0.85\columnwidth]{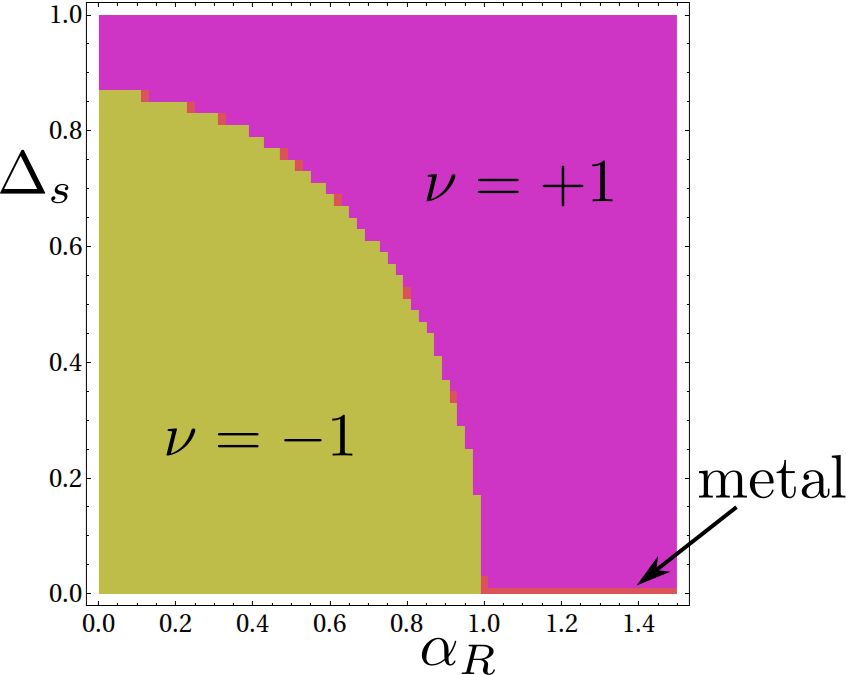}
\caption{(color online) Topological invariant $\nu$~as a function $\alpha_R$~and $\Delta_s$~at $\mu=0.5,~\Delta_p=1$. Green denotes the non-trivial phase ($\nu=-1$), purple denotes the trivial phase ($\nu=1$). The orange line at $\Delta_s=0,~\alpha_R>1$~indicates a metallic phase and the critical orange points at the phase boundary are also gapless.}
\label{fig:invariant}
\end{figure}

To show that our invariant Eq.~\eqref{eqn:nusimple} indeed distinguishes the topological from the trivial SCs in class DIII, we consider an example which does not exhibit any additional symmetry. Our model consists of two time-reversal copies of Kitaev's p-wave chain \cite{Kitaev2001}, coupled by a  Rashba spin-orbit term and augmented by an ordinary (s-wave) superconducting pairing term that competes with the p-wave coupling. The BdG Hamiltonian of this model reads
\begin{align}
H (k) =&
\bigl( 1 - \mu - \cos (k) \bigr) \sigma_0 \otimes \tau_z + 
 \Delta_p \sin (k) \, \sigma_0 \otimes \tau_y \nonumber \\
& +
\alpha_R \sin (k) \, \sigma_x \otimes \tau_z+\Delta_s\, \sigma_y \otimes \tau_y
\label{eq:modelham}
\end{align}
with $\mu$ the chemical potential, $\Delta_s, \Delta_p$ the SC pairings, $\alpha_R$ the Rashba spin-orbit coupling, where the energy is measured in units of the kinetic term. Recall that the $\sigma$ ($\tau$) Pauli matrices act in spin (particle-hole) space. For $\alpha_R = \Delta_s=0$, the system consists of two identical decoupled Kitaev chains. In Fig.~\ref{fig:invariant}, we show the $\alpha_R-\Delta_s$~phase diagram of this model for $\mu=0.5$, $\Delta_p = 1.0$. The data for Fig. \ref{fig:invariant} are obtained by direct evaluation of the topological invariant $\nu$~as defined in Eq. (\ref{eqn:nusimple}). We used a mesh of $n=1000$~points for the evaluation of Eq. (\ref{eqn:katoprop}) entering the definition of $\nu$. For $\Delta_s=0$~the gap closes for large spin orbit coupling $\alpha_R$~and a metallic phase emerges.\\

\section{Further simplification in the presence of inversion symmetry}
Even though we obtained a simple and numerically tractable form of the $\mathbb{Z}_2$ invariant, Eq.~\eqref{eqn:nusimple}, one can simplify the result even further in the presence of additional symmetries. We consider inversion symmetry, which has been used to simplify invariants in two- and three dimensional systems in class AII in \cite{FuInversion2007,HughesISTI2011}. Inversion symmetry is a symmetry of the model under $x \mapsto -x$, which in momentum space is implemented by the (momentum independent) unitary operator $P_{\rm inv}$, such that $P_{\rm inv} H(k) P_{\rm inv} = H(-k)$, with $P^2_{\rm inv} = 1$. We denote the eigenvalues of $P_{\rm inv}$ by $\xi_i = \pm 1$. We stress that we allow $P_{\rm inv}$ to also act non-trivially in spin and particle-hole space (apart from sending $x \mapsto -x$). 

The presence of inversion symmetry can generally be exploited in the following way \cite{FuInversion2007}.
In the first step, one shows that the Berry connection $\mathcal{A}_o (k)$ can be related to the
anti-symmetric, unitary matrix $pt(k)$, whose entries are the matrix elements of the operator $P_{\rm inv}\circ \mathcal{T}$. In particular, $\mathcal{A}_o (k) = -i \partial_k \log \bigl(\text{Pf}~pt_o (k)\bigr)$, where $pt_o(k)$ is the restriction of $pt(k)$ to the occupied bands \cite{FuInversion2007}. By adjusting the phase of the Bloch functions, one can set $\text{Pf}~pt_o(k) = 1$, implying that the Berry connection vanishes, and that the topological invariant can be obtained with knowledge about the system at the real momenta $k=0,\pi$ only. In the final step, one relates $\text{Pf}~\theta_o (0)$ and $\text{Pf}~ \theta_o (\pi)$ to the inversion symmetry eigenvalues $\xi_\alpha (0) = \pm 1$ and $\xi_\alpha (\pi) = \pm 1$ of the Kramers pairs. Because $[\mathcal{T},P_{\rm inv}] = 0$, both members of a Kramers pair have the same inversion symmetry eigenvalue. The final result is that $\text{Pf}~\theta_o (k) = \prod_{\alpha:\, {\rm occ}}' \xi_\alpha (k)$ for $k = 0,\pi$, where the product is over all occupied Kramers pairs, i.e., only one member of each pair contributes to the product. It follows that in the presence of inversion symmetry, one can write the invariant $\nu$, Eq.~\eqref{eqn:nusimple} in terms of the eigenvalues $\xi_i = \pm 1$ of $P_{\rm inv}$ at the real momenta as
\begin{equation}
\nu = \sideset{}{'}\prod_{\alpha:\, {\rm occ}} \xi_\alpha (0) \xi_\alpha (\pi) \ ,
\label{eq:nuinversion}
\end{equation}
where each occupied Kramers pair contributes once to the product. 

As a first application of this result we consider a generic 1DTSC in class D. For these non-TRS superconductors, Kitaev \cite{Kitaev2001} constructed a Pfaffian invariant $\mathcal{M} = \pm 1$. For translationally invariant systems, the invariant only involves the Pfaffian of the Majorana representation of the model at momenta $k = 0,\pi$. We recently showed that this invariant can be written in terms of the quantized Zak-Berry phase \cite{BudichDInv2013}. Although there are already several forms of the $\mathbb{Z}_2$ invariant available, it is interesting to note that in the presence of inversion symmetry, the invariant $\mathcal{M}$ can also be related to the eigenvalues of the operator $P_{\rm inv}$. The arguments given above for systems in class DIII rely on the presence of TRS. However, both members of each Kramers pair have the same inversion eigenvalue. It follows that if we `double' a 1D inversion symmetric superconductor with Hamiltonian $h$ in class D and construct a TRS model $H = \text{diag}\left(h,h^*\right)$, the $\mathbb{Z}_2$ invariant is given by Eq.~\eqref{eq:nuinversion}. The product can be taken over the occupied bands of the original system $h$. Hence, the invariant of inversion symmetric 1DTSC in class D is also given by Eq.~\eqref{eq:nuinversion}, but with the product running over all occupied bands (which are not Kramers degenerate).\\

\section{Concluding remarks}
We constructed a bulk topological invariant for time reversal symmetric superconductors in one dimension (corresponding to symmetry class DIII), which detects the presence or absence of a Kramers pair of Majorana bound states at the ends of the superconductor. The calculation of this invariant is numerically straightforward, because it does not require fixing of a phase relation between the Bloch states at different momenta. The only ingredients needed to calculate the invariant are the projections onto the occupied states, and the matrix elements of the TRS operator at the real momenta $k = 0,\pi$. We demonstrated our method by computing the topological phase diagram of a Rashba wire in the presence of two competing SC pairing terms, an s-wave and a p-wave pairing. For interacting systems, the BdG Hamiltonian can be replaced by the Nambu single particle Green's function $G$~ at zero frequency \cite{WangGeneralTOP,WangTSC,z2maj}, explicitly $H(k)\rightarrow -G^{-1}(\omega=0,k)$~in all calculations. 

In the presence of inversion symmetry, the topological invariant simplifies. It then only depends on the inversion symmetry eigenvalue associated with the Kramers pairs at the real momenta. Because the resulting invariant only depends on the parity associated with the Kramers pairs (both members share the same parity), one concludes that the same invariant can be used for 1D superconductors without TRS. Indeed, one can simply consider two time reversal conjugated copies of the same model. The same consideration holds for quantum anomalous Hall systems in 2D with inversion symmetry. Their Chern number can be calculated modulo two, by constructing two time reversal copies, and calculating the $\mathbb{Z}_2$ invariant associated with the resulting inversion symmetric quantum spin Hall system in class AII. The latter only depends on the inversion symmetry eigenvalues associated with the Kramers pairs at the time reversal invariant momenta \cite{FuInversion2007}.

{\em Acknowledgments}.
EA would like to thank R. Mong for interesting discussions. JCB would like to thank Patrik Recher for helpful comments.
This research was sponsored, in part, by the swedish research council.


\begin{thebibliography}{39}
\expandafter\ifx\csname natexlab\endcsname\relax\def\natexlab#1{#1}\fi
\expandafter\ifx\csname bibnamefont\endcsname\relax
  \def\bibnamefont#1{#1}\fi
\expandafter\ifx\csname bibfnamefont\endcsname\relax
  \def\bibfnamefont#1{#1}\fi
\expandafter\ifx\csname citenamefont\endcsname\relax
  \def\citenamefont#1{#1}\fi
\expandafter\ifx\csname url\endcsname\relax
  \def\url#1{\texttt{#1}}\fi
\expandafter\ifx\csname urlprefix\endcsname\relax\def\urlprefix{URL }\fi
\providecommand{\bibinfo}[2]{#2}
\providecommand{\eprint}[2][]{\url{#2}}

\bibitem[{\citenamefont{Kitaev}(2001)}]{Kitaev2001}
\bibinfo{author}{\bibfnamefont{A.}~\bibnamefont{Kitaev}},
  \bibinfo{journal}{Physics-Uspekhi} \textbf{\bibinfo{volume}{44}},
  \bibinfo{pages}{131} (\bibinfo{year}{2001}).

\bibitem[{\citenamefont{Hasan and Kane}(2010)}]{HasanKane}
\bibinfo{author}{\bibfnamefont{M.~Z.} \bibnamefont{Hasan}} \bibnamefont{and}
  \bibinfo{author}{\bibfnamefont{C.~L.} \bibnamefont{Kane}},
  \bibinfo{journal}{Rev. Mod. Phys.} \textbf{\bibinfo{volume}{82}},
  \bibinfo{pages}{3045} (\bibinfo{year}{2010}).

\bibitem[{\citenamefont{Qi and Zhang}(2011)}]{XLReview2010}
\bibinfo{author}{\bibfnamefont{X.-L.} \bibnamefont{Qi}} \bibnamefont{and}
  \bibinfo{author}{\bibfnamefont{S.-C.} \bibnamefont{Zhang}},
  \bibinfo{journal}{Rev. Mod. Phys.} \textbf{\bibinfo{volume}{83}},
  \bibinfo{pages}{1057} (\bibinfo{year}{2011}).

\bibitem[{\citenamefont{Beenakker}(2013)}]{BeenakkerMajReview}
\bibinfo{author}{\bibfnamefont{C.}~\bibnamefont{Beenakker}},
  \bibinfo{journal}{Annual Review of Condensed Matter Physics}
  \textbf{\bibinfo{volume}{4}}, \bibinfo{pages}{113} (\bibinfo{year}{2013}).

\bibitem[{\citenamefont{{Alicea}}(2012)}]{AliceaReview}
\bibinfo{author}{\bibfnamefont{J.}~\bibnamefont{{Alicea}}},
  \bibinfo{journal}{Reports on Progress in Physics}
  \textbf{\bibinfo{volume}{75}} (\bibinfo{year}{2012}).

\bibitem[{\citenamefont{Lutchyn et~al.}(2010)\citenamefont{Lutchyn, Sau, and
  Das~Sarma}}]{SauTSC}
\bibinfo{author}{\bibfnamefont{R.~M.} \bibnamefont{Lutchyn}},
  \bibinfo{author}{\bibfnamefont{J.~D.} \bibnamefont{Sau}}, \bibnamefont{and}
  \bibinfo{author}{\bibfnamefont{S.}~\bibnamefont{Das~Sarma}},
  \bibinfo{journal}{Phys. Rev. Lett.} \textbf{\bibinfo{volume}{105}},
  \bibinfo{pages}{077001} (\bibinfo{year}{2010}).

\bibitem[{\citenamefont{Oreg et~al.}(2010)\citenamefont{Oreg, Refael, and von
  Oppen}}]{OppenTSC}
\bibinfo{author}{\bibfnamefont{Y.}~\bibnamefont{Oreg}},
  \bibinfo{author}{\bibfnamefont{G.}~\bibnamefont{Refael}}, \bibnamefont{and}
  \bibinfo{author}{\bibfnamefont{F.}~\bibnamefont{von Oppen}},
  \bibinfo{journal}{Phys. Rev. Lett.} \textbf{\bibinfo{volume}{105}},
  \bibinfo{pages}{177002} (\bibinfo{year}{2010}).

\bibitem[{\citenamefont{Mourik et~al.}(2012)\citenamefont{Mourik, Zuo, Frolov,
  Plissard, Bakkers, and Kouwenhoven}}]{LeoMaj}
\bibinfo{author}{\bibfnamefont{V.}~\bibnamefont{Mourik}},
  \bibinfo{author}{\bibfnamefont{K.}~\bibnamefont{Zuo}},
  \bibinfo{author}{\bibfnamefont{S.~M.} \bibnamefont{Frolov}},
  \bibinfo{author}{\bibfnamefont{S.~R.} \bibnamefont{Plissard}},
  \bibinfo{author}{\bibfnamefont{E.~P. A.~M.} \bibnamefont{Bakkers}},
  \bibnamefont{and} \bibinfo{author}{\bibfnamefont{L.~P.}
  \bibnamefont{Kouwenhoven}}, \bibinfo{journal}{Science}
  \textbf{\bibinfo{volume}{336}}, \bibinfo{pages}{1003} (\bibinfo{year}{2012}).

\bibitem[{\citenamefont{{Deng} et~al.}(2012)\citenamefont{{Deng}, {Yu},
  {Huang}, {Larsson}, {Caroff}, and {Xu}}}]{LarssonXu}
\bibinfo{author}{\bibfnamefont{M.~T.} \bibnamefont{{Deng}}},
  \bibinfo{author}{\bibfnamefont{C.~L.} \bibnamefont{{Yu}}},
  \bibinfo{author}{\bibfnamefont{G.~Y.} \bibnamefont{{Huang}}},
  \bibinfo{author}{\bibfnamefont{M.}~\bibnamefont{{Larsson}}},
  \bibinfo{author}{\bibfnamefont{P.}~\bibnamefont{{Caroff}}}, \bibnamefont{and}
  \bibinfo{author}{\bibfnamefont{H.~Q.} \bibnamefont{{Xu}}},
  \bibinfo{journal}{Nano Lett.} \textbf{\bibinfo{volume}{12}},
  \bibinfo{pages}{6414} (\bibinfo{year}{2012}).

\bibitem[{\citenamefont{{Das} et~al.}(2012)\citenamefont{{Das}, {Ronen},
  {Most}, {Oreg}, {Heiblum}, and {Shtrikman}}}]{HeiblumMaj}
\bibinfo{author}{\bibfnamefont{A.}~\bibnamefont{{Das}}},
  \bibinfo{author}{\bibfnamefont{Y.}~\bibnamefont{{Ronen}}},
  \bibinfo{author}{\bibfnamefont{Y.}~\bibnamefont{{Most}}},
  \bibinfo{author}{\bibfnamefont{Y.}~\bibnamefont{{Oreg}}},
  \bibinfo{author}{\bibfnamefont{M.}~\bibnamefont{{Heiblum}}},
  \bibnamefont{and}
  \bibinfo{author}{\bibfnamefont{H.}~\bibnamefont{{Shtrikman}}},
  \bibinfo{journal}{Nature Phys.} \textbf{\bibinfo{volume}{8}},
  \bibinfo{pages}{887} (\bibinfo{year}{2012}).

\bibitem[{\citenamefont{{Alicea} et~al.}(2011)\citenamefont{{Alicea}, {Oreg},
  {Refael}, {von Oppen}, and {Fisher}}}]{AliceaBraiding}
\bibinfo{author}{\bibfnamefont{J.}~\bibnamefont{{Alicea}}},
  \bibinfo{author}{\bibfnamefont{Y.}~\bibnamefont{{Oreg}}},
  \bibinfo{author}{\bibfnamefont{G.}~\bibnamefont{{Refael}}},
  \bibinfo{author}{\bibfnamefont{F.}~\bibnamefont{{von Oppen}}},
  \bibnamefont{and} \bibinfo{author}{\bibfnamefont{M.~P.~A.}
  \bibnamefont{{Fisher}}}, \bibinfo{journal}{Nature Physics}
  \textbf{\bibinfo{volume}{7}}, \bibinfo{pages}{412} (\bibinfo{year}{2011}).

\bibitem[{\citenamefont{Altland and Zirnbauer}(1997)}]{AltlandZirnbauer}
\bibinfo{author}{\bibfnamefont{A.}~\bibnamefont{Altland}} \bibnamefont{and}
  \bibinfo{author}{\bibfnamefont{M.~R.} \bibnamefont{Zirnbauer}},
  \bibinfo{journal}{Phys. Rev. B} \textbf{\bibinfo{volume}{55}},
  \bibinfo{pages}{1142} (\bibinfo{year}{1997}).

\bibitem[{\citenamefont{{Roy}}(2008)}]{RoyTSC}
\bibinfo{author}{\bibfnamefont{R.}~\bibnamefont{{Roy}}},
  \bibinfo{journal}{arXiv:0803.2868}  (\bibinfo{year}{2008}).

\bibitem[{\citenamefont{Schnyder et~al.}(2008)\citenamefont{Schnyder, Ryu,
  Furusaki, and Ludwig}}]{Schnyder2008}
\bibinfo{author}{\bibfnamefont{A.~P.} \bibnamefont{Schnyder}},
  \bibinfo{author}{\bibfnamefont{S.}~\bibnamefont{Ryu}},
  \bibinfo{author}{\bibfnamefont{A.}~\bibnamefont{Furusaki}}, \bibnamefont{and}
  \bibinfo{author}{\bibfnamefont{A.~W.~W.} \bibnamefont{Ludwig}},
  \bibinfo{journal}{Phys. Rev. B} \textbf{\bibinfo{volume}{78}},
  \bibinfo{pages}{195125} (\bibinfo{year}{2008}).

\bibitem[{\citenamefont{Qi et~al.}(2009)\citenamefont{Qi, Hughes, Raghu, and
  Zhang}}]{QiTSC}
\bibinfo{author}{\bibfnamefont{X.-L.} \bibnamefont{Qi}},
  \bibinfo{author}{\bibfnamefont{T.~L.} \bibnamefont{Hughes}},
  \bibinfo{author}{\bibfnamefont{S.}~\bibnamefont{Raghu}}, \bibnamefont{and}
  \bibinfo{author}{\bibfnamefont{S.-C.} \bibnamefont{Zhang}},
  \bibinfo{journal}{Phys. Rev. Lett.} \textbf{\bibinfo{volume}{102}},
  \bibinfo{pages}{187001} (\bibinfo{year}{2009}).

\bibitem[{\citenamefont{{Liu} et~al.}(2013)\citenamefont{{Liu}, {Wong}, and
  {Law}}}]{LawD3Braiding}
\bibinfo{author}{\bibfnamefont{X.-J.} \bibnamefont{{Liu}}},
  \bibinfo{author}{\bibfnamefont{C.~L.~M.} \bibnamefont{{Wong}}},
  \bibnamefont{and} \bibinfo{author}{\bibfnamefont{K.~T.} \bibnamefont{{Law}}},
  \bibinfo{journal}{ArXiv e-prints}  (\bibinfo{year}{2013}),
  \eprint{1304.3765}.

\bibitem[{\citenamefont{Wong and Law}(2012)}]{Law1DD3}
\bibinfo{author}{\bibfnamefont{C.~L.~M.} \bibnamefont{Wong}} \bibnamefont{and}
  \bibinfo{author}{\bibfnamefont{K.~T.} \bibnamefont{Law}},
  \bibinfo{journal}{Phys. Rev. B} \textbf{\bibinfo{volume}{86}},
  \bibinfo{pages}{184516} (\bibinfo{year}{2012}).

\bibitem[{\citenamefont{Nakosai et~al.}(2013)\citenamefont{Nakosai, Budich,
  Tanaka, Trauzettel, and Nagaosa}}]{Nakosai1DD3}
\bibinfo{author}{\bibfnamefont{S.}~\bibnamefont{Nakosai}},
  \bibinfo{author}{\bibfnamefont{J.~C.} \bibnamefont{Budich}},
  \bibinfo{author}{\bibfnamefont{Y.}~\bibnamefont{Tanaka}},
  \bibinfo{author}{\bibfnamefont{B.}~\bibnamefont{Trauzettel}},
  \bibnamefont{and} \bibinfo{author}{\bibfnamefont{N.}~\bibnamefont{Nagaosa}},
  \bibinfo{journal}{Phys. Rev. Lett.} \textbf{\bibinfo{volume}{110}},
  \bibinfo{pages}{117002} (\bibinfo{year}{2013}).

\bibitem[{\citenamefont{Keselman et~al.}(2013)\citenamefont{Keselman, Fu,
  Stern, and Berg}}]{KeselmanDIII2013}
\bibinfo{author}{\bibfnamefont{A.}~\bibnamefont{Keselman}},
  \bibinfo{author}{\bibfnamefont{L.}~\bibnamefont{Fu}},
  \bibinfo{author}{\bibfnamefont{A.}~\bibnamefont{Stern}}, \bibnamefont{and}
  \bibinfo{author}{\bibfnamefont{E.}~\bibnamefont{Berg}},
  \bibinfo{journal}{arxiv:1305.4948}  (\bibinfo{year}{2013}).

\bibitem[{\citenamefont{Motrunich et~al.}(2001)\citenamefont{Motrunich, Damle,
  and Huse}}]{Motrunich2001}
\bibinfo{author}{\bibfnamefont{O.}~\bibnamefont{Motrunich}},
  \bibinfo{author}{\bibfnamefont{K.}~\bibnamefont{Damle}}, \bibnamefont{and}
  \bibinfo{author}{\bibfnamefont{D.~A.} \bibnamefont{Huse}},
  \bibinfo{journal}{Phys. Rev. B} \textbf{\bibinfo{volume}{63}},
  \bibinfo{pages}{224204} (\bibinfo{year}{2001}).

\bibitem[{\citenamefont{Nayak et~al.}(2008)\citenamefont{Nayak, Simon, Stern,
  Freedman, and Das~Sarma}}]{NayakReview}
\bibinfo{author}{\bibfnamefont{C.}~\bibnamefont{Nayak}},
  \bibinfo{author}{\bibfnamefont{S.~H.} \bibnamefont{Simon}},
  \bibinfo{author}{\bibfnamefont{A.}~\bibnamefont{Stern}},
  \bibinfo{author}{\bibfnamefont{M.}~\bibnamefont{Freedman}}, \bibnamefont{and}
  \bibinfo{author}{\bibfnamefont{S.}~\bibnamefont{Das~Sarma}},
  \bibinfo{journal}{Rev. Mod. Phys.} \textbf{\bibinfo{volume}{80}},
  \bibinfo{pages}{1083} (\bibinfo{year}{2008}).

\bibitem[{\citenamefont{Kitaev}(2009)}]{KitaevPeriodic}
\bibinfo{author}{\bibfnamefont{A.}~\bibnamefont{Kitaev}}, \bibinfo{journal}{AIP
  Conference Proceedings} \textbf{\bibinfo{volume}{1134}}, \bibinfo{pages}{22}
  (\bibinfo{year}{2009}).

\bibitem[{\citenamefont{{Ryu} et~al.}(2010)\citenamefont{{Ryu}, {Schnyder},
  {Furusaki}, and {Ludwig}}}]{RyuLudwig}
\bibinfo{author}{\bibfnamefont{S.}~\bibnamefont{{Ryu}}},
  \bibinfo{author}{\bibfnamefont{A.~P.} \bibnamefont{{Schnyder}}},
  \bibinfo{author}{\bibfnamefont{A.}~\bibnamefont{{Furusaki}}},
  \bibnamefont{and} \bibinfo{author}{\bibfnamefont{A.~W.~W.}
  \bibnamefont{{Ludwig}}}, \bibinfo{journal}{New Journal of Physics}
  \textbf{\bibinfo{volume}{12}}, \bibinfo{pages}{065010}
  (\bibinfo{year}{2010}).

\bibitem[{\citenamefont{{Fulga} et~al.}(2012)\citenamefont{{Fulga}, {Hassler},
  and {Akhmerov}}}]{FulgaSMatrixTI}
\bibinfo{author}{\bibfnamefont{I.~C.} \bibnamefont{{Fulga}}},
  \bibinfo{author}{\bibfnamefont{F.}~\bibnamefont{{Hassler}}},
  \bibnamefont{and} \bibinfo{author}{\bibfnamefont{A.~R.}
  \bibnamefont{{Akhmerov}}}, \bibinfo{journal}{\prb}
  \textbf{\bibinfo{volume}{85}}, \bibinfo{pages}{165409}
  (\bibinfo{year}{2012}).

\bibitem[{\citenamefont{Fu and Kane}(2007)}]{FuInversion2007}
\bibinfo{author}{\bibfnamefont{L.}~\bibnamefont{Fu}} \bibnamefont{and}
  \bibinfo{author}{\bibfnamefont{C.~L.} \bibnamefont{Kane}},
  \bibinfo{journal}{Phys. Rev. B} \textbf{\bibinfo{volume}{76}},
  \bibinfo{pages}{045302} (\bibinfo{year}{2007}).

\bibitem[{\citenamefont{Turner et~al.}(2011)\citenamefont{Turner, Pollmann, and
  Berg}}]{TurnerEnt2011}
\bibinfo{author}{\bibfnamefont{A.~M.} \bibnamefont{Turner}},
  \bibinfo{author}{\bibfnamefont{F.}~\bibnamefont{Pollmann}}, \bibnamefont{and}
  \bibinfo{author}{\bibfnamefont{E.}~\bibnamefont{Berg}},
  \bibinfo{journal}{Phys. Rev. B} \textbf{\bibinfo{volume}{83}},
  \bibinfo{pages}{075102} (\bibinfo{year}{2011}).

\bibitem[{\citenamefont{Fidkowski and Kitaev}(2011)}]{Fidkowski2011}
\bibinfo{author}{\bibfnamefont{L.}~\bibnamefont{Fidkowski}} \bibnamefont{and}
  \bibinfo{author}{\bibfnamefont{A.}~\bibnamefont{Kitaev}},
  \bibinfo{journal}{Phys. Rev. B} \textbf{\bibinfo{volume}{83}},
  \bibinfo{pages}{075103} (\bibinfo{year}{2011}).

\bibitem[{\citenamefont{Berry}(1984)}]{Berry}
\bibinfo{author}{\bibfnamefont{M.~V.} \bibnamefont{Berry}},
  \bibinfo{journal}{Proc. R. Soc. Lond. A} \textbf{\bibinfo{volume}{392}},
  \bibinfo{pages}{45} (\bibinfo{year}{1984}).

\bibitem[{\citenamefont{Zak}(1989)}]{ZakPol}
\bibinfo{author}{\bibfnamefont{J.}~\bibnamefont{Zak}}, \bibinfo{journal}{Phys.
  Rev. Lett.} \textbf{\bibinfo{volume}{62}}, \bibinfo{pages}{2747}
  (\bibinfo{year}{1989}).

\bibitem[{\citenamefont{Hatsugai}(2006)}]{HatsugaiQuantizedBerry}
\bibinfo{author}{\bibfnamefont{Y.}~\bibnamefont{Hatsugai}},
  \bibinfo{journal}{Journal of the Physical Society of Japan}
  \textbf{\bibinfo{volume}{75}}, \bibinfo{pages}{123601}
  (\bibinfo{year}{2006}).

\bibitem[{\citenamefont{Fu and Kane}(2006)}]{FuPump}
\bibinfo{author}{\bibfnamefont{L.}~\bibnamefont{Fu}} \bibnamefont{and}
  \bibinfo{author}{\bibfnamefont{C.~L.} \bibnamefont{Kane}},
  \bibinfo{journal}{Phys. Rev. B} \textbf{\bibinfo{volume}{74}},
  \bibinfo{pages}{195312} (\bibinfo{year}{2006}).

\bibitem[{\citenamefont{Kato}(1950)}]{Kato1950}
\bibinfo{author}{\bibfnamefont{T.}~\bibnamefont{Kato}},
  \bibinfo{journal}{Journal of the Physical Society of Japan}
  \textbf{\bibinfo{volume}{5}}, \bibinfo{pages}{435} (\bibinfo{year}{1950}).

\bibitem[{\citenamefont{Avron et~al.}(1989)\citenamefont{Avron, Sadun, Segert,
  and Simon}}]{AvronQuadrupole1989}
\bibinfo{author}{\bibfnamefont{J.~E.} \bibnamefont{Avron}},
  \bibinfo{author}{\bibfnamefont{L.}~\bibnamefont{Sadun}},
  \bibinfo{author}{\bibfnamefont{J.}~\bibnamefont{Segert}}, \bibnamefont{and}
  \bibinfo{author}{\bibfnamefont{B.}~\bibnamefont{Simon}},
  \bibinfo{journal}{Communications in Mathematical Physics}
  \textbf{\bibinfo{volume}{124}}, \bibinfo{pages}{595} (\bibinfo{year}{1989}).

\bibitem[{\citenamefont{{Budich} and {Trauzettel}}(2013)}]{TSMReview}
\bibinfo{author}{\bibfnamefont{J.~C.} \bibnamefont{{Budich}}} \bibnamefont{and}
  \bibinfo{author}{\bibfnamefont{B.}~\bibnamefont{{Trauzettel}}},
  \bibinfo{journal}{Physica Status Solidi Rapid Research Letters}
  \textbf{\bibinfo{volume}{7}}, \bibinfo{pages}{109} (\bibinfo{year}{2013}).

\bibitem[{\citenamefont{Hughes et~al.}(2011)\citenamefont{Hughes, Prodan, and
  Bernevig}}]{HughesISTI2011}
\bibinfo{author}{\bibfnamefont{T.~L.} \bibnamefont{Hughes}},
  \bibinfo{author}{\bibfnamefont{E.}~\bibnamefont{Prodan}}, \bibnamefont{and}
  \bibinfo{author}{\bibfnamefont{B.~A.} \bibnamefont{Bernevig}},
  \bibinfo{journal}{Phys. Rev. B} \textbf{\bibinfo{volume}{83}},
  \bibinfo{pages}{245132} (\bibinfo{year}{2011}).

\bibitem[{\citenamefont{Budich and Ardonne}(2013)}]{BudichDInv2013}
\bibinfo{author}{\bibfnamefont{J.~C.} \bibnamefont{Budich}} \bibnamefont{and}
  \bibinfo{author}{\bibfnamefont{E.}~\bibnamefont{Ardonne}},
  \bibinfo{journal}{arXiv:1306.4459}  (\bibinfo{year}{2013}).

\bibitem[{\citenamefont{Wang and Zhang}(2012{\natexlab{a}})}]{WangGeneralTOP}
\bibinfo{author}{\bibfnamefont{Z.}~\bibnamefont{Wang}} \bibnamefont{and}
  \bibinfo{author}{\bibfnamefont{S.-C.} \bibnamefont{Zhang}},
  \bibinfo{journal}{Phys. Rev. X} \textbf{\bibinfo{volume}{2}},
  \bibinfo{pages}{031008} (\bibinfo{year}{2012}{\natexlab{a}}).

\bibitem[{\citenamefont{Wang and Zhang}(2012{\natexlab{b}})}]{WangTSC}
\bibinfo{author}{\bibfnamefont{Z.}~\bibnamefont{Wang}} \bibnamefont{and}
  \bibinfo{author}{\bibfnamefont{S.-C.} \bibnamefont{Zhang}},
  \bibinfo{journal}{Phys. Rev. B} \textbf{\bibinfo{volume}{86}},
  \bibinfo{pages}{165116} (\bibinfo{year}{2012}{\natexlab{b}}).

\bibitem[{\citenamefont{Budich and Trauzettel}(2013)}]{z2maj}
\bibinfo{author}{\bibfnamefont{J.~C.} \bibnamefont{Budich}} \bibnamefont{and}
  \bibinfo{author}{\bibfnamefont{B.}~\bibnamefont{Trauzettel}},
  \bibinfo{journal}{New Journal of Physics} \textbf{\bibinfo{volume}{15}},
  \bibinfo{pages}{065006} (\bibinfo{year}{2013}).

\end{thebibliography}

\end{document}